\documentstyle[aps,12pt,prc,epsf]{revtex}
\newcommand{\ve}{\varepsilon}
\newcommand{\be}{\begin{equation}}
\newcommand{\ee}{\end{equation}}
\draft
\tighten

\title{Quantum-Mechanical Analysis of Single Particle Level 
  Density\footnote{
  Contribution to {\it European (Int.) Conference on Advances in 
  Nuclear Physics and Related Areas, Thessaloniki, Greece, 8-12 July 
  1997}; Rom. J. Phys. {\it (in press)}.  
}} 
\author{I. \c Ste\c tcu\footnote{stetcu@roifa.ifa.ro} } 
\address{Institute for Nuclear Physics and Engineering "Horia Hulubei",\\
  P.O. Box MG-6, 76900 Bucharest, Romania}

\begin{document}
\maketitle

\begin{abstract}
A quantum-mechanical calculation of the single-particle level (s.p.l.) 
density $g(\varepsilon)$ is carried on by using the connection with 
the single-particle Green's function. The relation between the 
imaginary part of Green's function and single-particle wave functions 
is used separately for the discrete and continuous states. Within the 
bound-states region the imaginary part of the Green's function is 
calculated by using the wronskian theorem. The Green's function 
corresponding to the continuum is written by using the regular 
and Jost solutions of the radial Schr\"odinger equation. The smooth 
part of the rapidly fluctuating s.p.l. density is calculated by means 
of the Strutinski procedure. The continuum component of the s.p.l. 
density has rather close values within either exact quantum-mechanical  
calculations with the Woods-Saxon (WS) potential, or Thomas-Fermi 
approximation with WS as well as finite-square potential wells, 
provided that the free-gas contribution is subtracted. A similar 
trend is obtained by means of the simple FGM formula for the s.p.l. 
density if the continuum effect is taken into account.
\end{abstract}
\section{Introduction}

The nuclear level density has a central role in the statistical 
analysis of nuclear reactions. Information concerning nuclear 
densities can be obtained from experiments by means of the analysis
of neutron and charged particles resonances, inelastic scattering, 
and particle evaporation spectra, so that the theoretical 
calculation of this quantity is quite useful for validation of nuclear 
models.
Many approaches (e.g. \cite{shl92,shl96,shl97} and references therein) 
have been developed to calculate the single-particle level (s.p.l.) 
density on which is based the nuclear level density calculation, its 
establishment being yet considered a difficult task. A constant s.p.l. 
density $g \sim A/14$ MeV$^{-1}$ was used in various statistical model 
descriptions of the nuclear reactions. An increased criticism has been 
expressed on this constant value which is inconsistent with the number 
of $A$ nucleons in the nucleus and the usual Fermi energy $F=38$MeV.

The semiclassical approximations could be the starting point of a basic
analysis of the s.p.l. density, including the energy dependence. First,
the Thomas-Fermi approximation proves adequate for a  special class of 
potential wells \cite{shl92} which describe the nuclear mean field 
(Woods-Saxon, harmonic oscillator, trapezoidal potential well). However,
for potentials like square-well and infinite square potential, 
the results deviate significantly from the quantum-mechanical 
calculation. Enhanced approximations add the lowest terms in $\hbar$ 
as corrections to the Thomas-Fermi results \cite{shl92,shl96}. 

An exact quantum-mechanical calculation was provided by Shlomo 
\cite{shl92} by using a Green's function approach. A particular case 
is the s.p.l. density for a finite potential well. Since it is 
including the free-gas states, one can  calculate and subtract this 
component by using the Green's function associated with the respective 
single-particle Hamiltonian. Shlomo found by means of both 
semiclassical and quantum methods that, for a realistic finite depth 
potential well, the s.p.l. density {\it decreases} with energy in the 
continuum region ({\it the continuum effect}).

The paper is organised as follows. Section II is an introduction of 
the general theoretical aspects concerning the s.p.l. density, namely 
the definition, its relation with the Green's function, the Strutinski 
smoothing procedure, and semiclassical methods usually involved. 
Section III gives a general description of the single particle Green's 
function theory and a method for the calculation of its imaginary 
part. The model parameters for the mean field potential and numerical 
results are discussed in section IV. Finally, the conclusions are 
drawn in section V.

\section{Basic Formalism for Single-Particle Level Density }

\subsection{The quantum-mechanical single-particle level density }

A brief presentation  of the  s.p.l. density definition, Green's 
function and relation between them are given in this section. The 
s.p.l. density is defined as \cite{shl92}

\begin{equation} \label{1.1}
g(\varepsilon)=\mbox{Tr}(\delta (\varepsilon-\hat H)) \: ,
\end{equation}
where $\hat H$ is the corresponding single-particle Hamiltonian (mean
field)

\begin{equation}
\label{1.2}
\hat H=\frac{{\bf \hat P}^2}{2m}+V(r) \: .
\end{equation}

The eventual bound states of the Hamiltonian given by Eq. (\ref{1.2})  
can be obtained from the time independent Schr\"odinger equation
$$
\hat H|n\alpha \rangle =\varepsilon_n|n\alpha \rangle \: ,
$$
$\alpha $ being the degeneracy given by other quantum numbers. It is
supposed that the normalisation for the eigenvectors corresponding to 
bound states is

\begin{equation} \label{1.4}
\langle n^{\prime}\alpha^{\prime}|n \alpha
\rangle=\delta_{nn^{\prime}}\delta_{\alpha\alpha^{\prime}},
\end{equation}
so that the s.p.l. density following Eq. (\ref{1.1}) is

\begin{equation} \label{1.11}
g_B(\varepsilon)=\sum_{n\alpha}\delta(\varepsilon-\varepsilon_n) \: .  
\end{equation}

In the case of a finite potential well which has bound as well as 
continuous states the s.p.l. density can be considered for the two 
regions of the spectrum

\[
g(\ve)=g_B(\ve)+g_C(\ve) \: ,
\]
where $g_B(\ve)$ and $g_C(\ve)$ are the contributions given by the
bound and continuus states respectively. The continuum contribution 
are determined from the scattering phase shifts 
$\delta _{lj}(\varepsilon)$ by \cite{shl92,ross}

$$
g_C(\varepsilon)=\frac{1}{\pi}\sum_{lj}(2j+1)\frac{\mbox{d}\delta_{lj}
(\varepsilon)}{\mbox{d}\varepsilon} \: . 
$$
It can be determined also from Eq. (\ref{1.1}) but the eigenvectors 
corresponding to the continuum states can not be normalised similarly 
to the bound states case. The relation (\ref{1.4}) becoming

\begin{equation} \label{1.5}
\langle \varepsilon^{\prime}\alpha^{\prime}|\varepsilon \alpha
\rangle=\delta(\varepsilon-\varepsilon')\delta(\alpha-\alpha') \: ,  
\end{equation}
it results

\begin{equation} \label{1.6}
g_C(\varepsilon)=\sum_{\alpha ^{\prime }}\int \mbox{d}\varepsilon'
\delta(\varepsilon-\varepsilon^{\prime })\langle \varepsilon'\alpha'
|\varepsilon^{\prime }\alpha^{\prime }\rangle \: ,
\end{equation}
where the sum after $\alpha ^{\prime }$ can be an integral if this 
parameter is continuous. For the moment we will assume that the 
states are not degenerate so that the s.p.l. density in the whole 
energy range has the form

\begin{equation} \label{1.7}
g(\varepsilon)=\sum_n \delta(\varepsilon-\varepsilon_n)+  \int 
\mbox{d}\varepsilon^{\prime}\delta(\varepsilon-\varepsilon')\langle
\varepsilon^{\prime}|\varepsilon^{\prime}\rangle \:.  
\end{equation}

An important quantity is the number of states $N(\ve)$ with energy 
bellow  $\ve$, which will be used latter for illustration of the 
{\it continuum effect},

\be
\label{num}
N(\ve)=\int_{-\infty}^\ve \mbox{d} \ve' g(\ve') \: .
\ee
By means of the condition
\[
A=\int_{-\infty}^{\ve_F}d\ve g(\ve) \:,
\] 
Eq. (\ref{num}) can be also used for determination of the Fermi energy
$\ve_F$ which is a significant quantity for the description
of the nucleus ground state or nuclear reactions analysis.

\subsection{The Green's function approach }

The single-particle Green's function $G({\bf {r},{r}';\varepsilon)}$ 
corresponding to the Hamiltonian (\ref{1.2}) is defined by the 
equation \cite{shl92,shl97,new66}

\begin{equation} \label{1.8}
(\varepsilon-\hat H)G({\bf r,r'};\varepsilon)=\delta({\bf r-r'}) \: .
\end{equation}

A first assumption has considered the nucleus in a large volume 
$\Omega$ ({\it i.e.} within an infinite potential well), the continuum 
being discretized. In this case, the spectral representation of the 
Green's function

\begin{equation} \label{1.9}
  G({\bf r,r'};\varepsilon)=\sum_n \frac{ \langle {\bf r}|n
  \rangle \langle n |{\bf r'}\rangle}{\varepsilon-\varepsilon_n}
\end{equation}
can be used to obtain

\begin{eqnarray} \label {1.10}
 & &\lim_{\alpha\to 0_+}\frac{1}{2\pi i}\int\mbox{d}{\bf r} 
    \lim_{{\bf r}'\to{\bf r}}\left(G({\bf r, r}';\varepsilon+i\alpha)-
                G({\bf r, r}';\varepsilon-i\alpha)\right) \nonumber \\
 &=& \frac{1}{2\pi i}\lim_{\alpha\to 0+}\sum_n
    \left(\frac{1}{\varepsilon-\varepsilon_n+i\alpha}-
    \frac{1}{\varepsilon-\varepsilon_n-i\alpha}\right)
    =-\sum_n\delta(\varepsilon-\varepsilon_n) \: ,
\end{eqnarray}
where it has been used the well-known relation \cite{new66}

$$
  \lim_{\alpha\to 0_+}\frac{1}{x-a\pm i\alpha}=
                      \frac{P}{x-a}\mp i\pi \delta(x-a) \: .
$$
By comparison of the Eqs. (\ref{1.11}) and (\ref{1.10}) it results the 
following relation between the s.p.l. density and Green's function
(see also \cite{goriely96})

\begin{equation} \label{1.12}
g(\varepsilon)=-\frac{1}{\pi} \lim_{\alpha\to 0_+}\int\mbox{d}{\bf r}
   \lim_{{\bf r \to r}^{\prime}} \mbox{Im} 
   (G({\bf r, r}^{\prime};\varepsilon+i\alpha)) \: .
\end{equation} 

The demonstration can be generalised at this point for the both sides
of the spectrum. Without consideration of the nucleus in a finite 
volume Eq. (\ref{1.9}) becomes

\begin{equation} \label{1.91}
  G({\bf r,r'};\varepsilon)=\sum_n \frac{ \langle {\bf r}| n
  \rangle \langle n |{\bf r'}\rangle}{\varepsilon-\varepsilon_n}+
  \int \mbox{d}\ve' \frac{ \langle {\bf r}| \ve' \rangle 
  \langle \ve' |{\bf r'}\rangle}{\varepsilon-\varepsilon'}
\end{equation}
while an integral corresponding to the continuum energy range is added
similarly to the sum in Eq. (\ref{1.10}). Next, within this additional 
term it can be used the relation

$$
  \int \mbox{d} {\bf r} \langle{ \bf r}|\varepsilon'\rangle\langle
  \varepsilon'| {\bf r} \rangle= 
  \langle\varepsilon'|\varepsilon'\rangle \: ,
$$
so that the respective continuum contribution becomes just the second 
term in Eq. (\ref{1.7}). Therefore, Eq. (\ref{1.12}) is proved worthy
not only for the bound states but in the general case.

Next, the dependence of the s.p.l. density by $\ve$ and ${\bf r}$ can 
be expressed as

\be \label{1.13}
\left. g(\ve,{\bf r})=-\frac{1}{\pi}\lim_{\alpha\to 0_+} \mbox{Im}
     G({\bf r,r}';\ve+i\alpha)\right|_{{\bf r \to r}'} \: .
\ee
It is reduced to the dependence by $\ve$ and $r$ in the common 
case of a central potential.

The s.p.l. density for the free particles without spin, {\it i.e.} for
$V(r)=0$ in Eq. (\ref{1.2}) can be also obtained for Eqs. (\ref{1.12})
and (\ref{1.13}). The single particle Green's function in this case 
is

\be \label{1.14}
 G^{(+)}_0({\bf r,r}';\ve)=\lim_{\alpha\to 0_+}G_0({\bf r, r}',\ve+
  i\alpha)=-\frac{2m}{\hbar^2}\frac{\exp(ik|{\bf R}|)}
  {4\pi |{\bf R}|} \; ,
\ee
where ${\bf R=r-r'}$ and $k=\sqrt{2m\ve/\hbar^2}$.

Introducing Eq. (\ref{1.14}) in Eq. (\ref{1.13})

\begin{eqnarray}
g_0(\ve,{\bf r})&=&-\frac{1}{\pi}\lim_{R \to 0} \mbox{Im} 
 \left[-\frac{2m}{\hbar^2}\frac{\exp(ikR)}{4\pi R}\right] \nonumber \\
 &=& \frac{2m}{4\pi^2\hbar^2}
     \lim_{R \to 0}\left[ \frac{\sin(kR)}{R} \right],
\end{eqnarray}
it results the expected characteristic of $g_0(\ve,{\bf r})$ to be a 
function of only $\ve$, for the case of free particles

\be \label{1.15}
   g_0(\ve,{\bf r})=\frac{1}{4\pi^2} 
       \left( \frac{2m}{\hbar^2} \right)^{3/2} \sqrt{\ve} \: .
\ee

Furthermore, by using Eq. (\ref{1.15}) in Eq. (\ref{1.12}) one obtains 
the well-known formula of the s.p.l. density for free particles 
without spin restricted to a finite volume $\Omega$ \cite{shl92}

\be \label{1.16}
   g_0(\ve)=\frac{\Omega}{4\pi^2} 
       \left( \frac{2m}{\hbar^2} \right)^{3/2} \sqrt{\ve} \: .
\ee 

We have to pay attention at the numerical calculation of the limit
$\alpha \to 0_+$ in Eq. (\ref{1.12}). The s.p.l. density could be 
calculated by smearing of the Green's function with a Lorentz function 
of small width $\gamma$. By using the relation

\be \label{1.17}
\lim_{\alpha \to 0_+} \frac{\gamma}{\pi}\int_{-\infty}^\infty
  \frac{\mbox{d}\ve'}{(\ve'-\ve)^2+\gamma^2}G({\bf r,r}';\ve'+i\alpha)
  =G({\bf r,r}';\ve+i\gamma) \: ,
\ee
the smeared level density $g_\gamma(\ve)$ can be obtained from Eq.
(\ref{1.12}) while $\alpha$ is replaced by a finite but small value 
of $\gamma$. This parameter should have a value much smaller than 
$\hbar\omega$, the spacing between major shells. The method is used 
in several studies \cite{shl92,shl97,bal70} and has the advantage that 
the knowledge of the spectrum of $\hat H$ is not any more necessary.

Alternatively, an analytical calculation of the single-particle 
Green's function is adopted in the present work, in order to obtain 
an exact form of the s.p.l. density within the continuum region too. 
Because of the differences between the regular and irregular solutions 
of the radial Schr\"odinger equation for the two regions of the 
spectrum, the respective limits $\alpha\to 0_{+}$ in Eq. (\ref{1.13}) 
have to be calculated separately as it is shown in section III. 

\subsection{The smoothing procedure }

The level density $g(\ve)$ varies rapidly with energy, due to the 
shell effects. However, besides its rapidly fluctuating part, $g(\ve)$ 
contains a smooth part describing the average behaviour of the level 
distribution \cite{ross,bra}

\[
   g(\ve)=g_s(\ve)+\delta g(\ve) \: .
\]

For the discrete states, the smooth part of the level density can 
be obtained from Eq. (\ref{1.12}) by smearing out each $\delta$ 
function to a Gaussian of width $\Gamma$ \cite{ross}

\[
  \delta(\ve-\ve_n)\to \frac{1}{\Gamma\sqrt{\pi}}
   \exp\left[-\left(\frac{\ve-\ve_n}{\Gamma}\right)^2\right] \: ,
\]
where $\Gamma$ should be at least of the order of the shell spacing. 
Although such a Gaussian preserves the normalisation of a 
$\delta$-function, curvature corrections have to be introduced in such 
a way that not only $g_s(\ve)$ but also its first $2M$ derivatives are 
reproduced correctly \cite{ross,bra}. The unitary treatment of the 
bound and continuous states is realized by means of an average of the 
s.p.l. density (Strutinski's procedure) \cite{ross}

\be \label{1.20}
   g_s(\ve)=\int_{-\infty}^\infty \mbox{d}\ve' F(\ve-\ve')g(\ve') \: ,
\ee
where 
\[
 F(x)=\frac{1}{\Gamma\sqrt{\pi}}\mbox{e}^{
 -\left(\frac{x}{\Gamma}\right)^2}L^{\frac{1}{2}}_M(x^2/\Gamma^2) \: ,
\]
and $L^{\frac{1}{2}}_M$ is the associated Laguerre polynomial 
\cite{abr}. The smoothing parameter $\Gamma$ has to be chosen 
\cite{ross,bra} greater than $\hbar\omega$.

\subsection{Subtracting the free gas contribution }

The background level density due to the free Fermi gas (which goes 
like $\sqrt{\ve}$) exists whether or not the potential well is finite 
\cite{ross}, the former case yielding the quantum fluctuation of the 
density. A subtracting procedure was introduced by Shlomo 
\cite{shl92,shl96,shl97}, the states corresponding to the free 
Hamiltonian $\hat H={\bf \hat P}/2m$ being taken away for $\ve > 0$

\be \label{subs}
   g(\ve)=\int \mbox{d}{\bf r} 
          \left(g(\ve,{\bf r})-g_0(\ve,{\bf r})\right) \: .
\ee
Therefore, the s.p.l. density decreases with energy in the continuum 
region (the so called {\it continuum effect}).

A particular problem occurs for protons. The corresponding free gas 
contribution has to be considered with the Coulomb interaction
included. The smoothing procedure (\ref{1.20}) will be involved 
following the subtraction of the free gas contribution.

\subsection{Semiclassical calculation of $g_s(\ve)$ }

The semiclassical approximations for calculation of the s.p.l. density 
are commonly used \cite{shl92,shl96,shl97,ring}. The $\hbar$-expansion 
(Wigner-Kirkwood) of the single-particle partition function is used in 
this respect \cite{ring}, performing the so-called Wigner transform.
The terms with odd power of $\hbar$ vanish for the smooth 
single-particle potentials. Therefore, the Thomas-Fermi (TF) 
approximation ($\hbar^0$ term) and the semiclassical (SC) 
approximation (TF+$\hbar^2$ correction) are generally considered. The 
numerical calculations in this work have been carried out for the 
TF-method only. The results of the TF approximation by taking into 
account the spin degeneracy and neglecting the spin-orbit interactions 
are given in several papers \cite{shl92,shl96,ring}

\be \label{tf}
  g_{TF}(\ve)=\frac{1}{2\pi}\left(\frac{2m}{\hbar}^2\right)^{3/2}
              \int \mbox{d}{\bf r}\left(\ve-V({\bf r})\right)^{1/2}
              \Theta(\ve-V({\bf r})) \: .
\ee
For the finite well potentials we have to subtract \cite{shl92,shl96}
the contribution of free Fermi gas when $\ve>0$, which is given by 
Eq. (\ref{1.15}).

\section{Single-Particle Green's Function Calculation }

In this section a partial wave analysis for Green's function is 
presented for particles with spin \cite{new66}. A brief review of some 
elements which are necessary for such description of particles is 
given firstly.

Let $\chi^s_\nu$ and $C(lsj,m\nu M)$ be the normalised eigenvectors of 
the spin, and respectively the Clebsch-Gordan coefficients. By using 
spherical harmonics $Y_l^m({\bf \hat r})$ there are defined the 
functions \cite{new66}

\be \label{1.21}
   {\bf Y}_{jls}^M({\bf \hat r})=
    \sum_{m\nu }C(lsj,m\nu M)Y_l^m({\bf \hat r}) \chi _\nu ^s \: .
\ee
They are eigenfunctions of the total angular-momentum operator, its 
$z$ component (with respect to a given $z$ axis), the orbital 
angle-momentum operator, and  the spin operator.

Following the expansion of the Green's function \cite{new66}

\be \label{1.22}
  G^{(\pm)}({\bf r,r^{\prime }};\ve)=
  \frac{2m}{\hbar ^2}\frac 1{rr^{\prime}}\sum_{jMls}
  {\bf Y}_{jls}^M({\bf \hat r}){\bf Y}_{jls}^{M*}({\bf \hat r^{\prime}})
  G_{lj}^{(\pm)}(r,r^{\prime };\ve) \: , 
\ee
the partial-wave Green's function $G_{lj}^{(\pm)}(r,r^{\prime };\ve)$ 
satisfies the equation

\be \label{1.18}
   \left(-\frac{\partial ^2}{\partial r^2}+\frac{l(l+1)}{r^2}-k^2+
   \frac{2m}{\hbar^2}V_{lj}(r)\right)G_{lj}^{(\pm)}(r,r^{\prime };\ve)
   =-\delta (r-r') 
\ee
with the boundary conditions specifying that it is zero at $r=0$, for 
fixed $r'$ and at $r\to\infty$ it contains outgoing (for $G^{(+)}$) 
or incoming waves (for $G^{(-)}$). The solution of Eq. (\ref{1.18}) 
can be expressed by using the regular $u_{lj}(r)$ and irregular 
solutions $v_{lj}^{(\pm)}(r)$ of the radial equation,

\be \label{21.3}
  G^{(\pm)}_{lj}(r,r';\ve)=
		\frac{u_{lj}(r_<)v_{lj}^{(\pm)}(r_>)}{W_{lj}} \: ,
\ee
where $W_{lj}$ is the wronskian of the two solutions, and $r_<$ and 
$r_>$ are the lesser and the greater of $r$ and $r'$, respectively. 
The regular solutions are defined for $r\to 0$ so that it is 
satisfied the boundary condition 

\be \label{21.1}
    u_{lj}(r=0)= 0 \: .
\ee
Because of the hermiticity of the Hamiltonian, the regular solution 
can be chosen real. The irregular solutions are defined for 
$r \to \infty$ by

\be \label{21.2}
    \left.v_{lj}^{(\pm)}(r)\right|_{r\to \infty} \sim \left\{ 
      \begin{array}{lcl} 
      \mbox{e}^{-\kappa r} & \mbox{if} & \ve <0 \\
      \\
      \mbox{e}^{\pm ikr}   & \mbox{if} & \ve >0 \: , \\
      \end{array}\right. 
\ee
where $\kappa=\sqrt{-2m\ve/\hbar^2}$.

\subsection{Bound-states region ($\ve<0$)}

The imaginary part of $G(r,r;\ve+i\alpha)$ is obtained following the
calculation of the limit $\alpha\to 0_{+}$. Moreover, we have to pay 
attention to the 
both possibilities for $\ve$ to be or not an eigenvalue. In the first 
case, since the regular and irregular solutions can be chosen real 
and are linearly independent, the limit in Eq. (\ref{1.12}) is easy 
to perform and the imaginary part of Green's function becomes zero.

In the second case the regular $u(r)$ and irregular solution $v(r)$
are linearly dependent, so that it results

\be \label{21.4}
    u_{lj}(r)=C_{lj}v_{lj}(r) \: ,
\ee
where the coefficient $C_{lj}$ can be obtained from the boundary 
conditions.

Let $\ve_0$ be an eigenvalue, and $K_0$ a complex number given by

\[ 
  K_0=\sqrt{\frac{2m(\ve_0+i\alpha)}{\hbar^2}}=i\kappa_0-\alpha' \: ,
\]
where $\alpha'\to 0_+$. The wronskian can be expanded around $\alpha'$

\be \label{21.5}
    W_0(K_0)=-\left.\frac{\partial W_0}{\partial K_0}
                    \right|_{K_0=i\kappa_0}\alpha'+O(\alpha'^2) \: ,
\ee
while the use of the wronskian theorem leads to \cite{new66,joac}

\be \label{21.6}
    \frac{\partial W_0}{\partial K_0}=-\frac{K_0}{C_0}
                        \int_0^\infty\mbox{d}ru^2_{lj}(r) \: .
\ee

The imaginary part of the Green's function can be calculated now by 
using Eqs. (\ref{21.3}) and (\ref{21.6}). The result is a sum of 
$\delta$-functions

\be \label{21.7}
    \mbox{Im}\left(G(r,r;\ve)\right)=
                     \sum_{n}D_{n}u^2_{n}(r)\delta(\ve-\ve_{n}) \: .
\ee
Finally, by introducing Eq. (\ref{21.7}) in Eq. (\ref{1.12}) and 
comparing with Eq. (\ref{1.11}), the coefficients $D_n$ can be 
obtained and the s.p.l. density at negative energies becomes

\be \label{21.8}
    g(\ve,r)=\sum_{n}u_n^2(r)\delta(\ve-\ve_n) \: ,
\ee
where it has been assumed the following normalisation condition for 
the regular solutions

\[
   \int_0^\infty \mbox{d}r u_n^2(r)=1 \: .
\]

The solutions of the radial Schr\" odinger equation considered in this
frame

\be \label{sch}
   \frac{\hbar^2}{2m}\frac{d^2u_n(r)}{dr^2}+\left[ \ve_n-V(r)-
   \frac{\hbar^2}{2m}\frac{l(l+1)}{r^2}\right]u_n(r)=0 \: ,
\ee
have been obtained for $V(r)$ being the sum of the Woods-Saxon (WS)
nuclear potential 
and the Coulomb potential in the case of protons. The former has the
expression

\be \label{ws}
     V^{WS}(r)=\frac{V_0}{1+\exp[(r-R)/a_0]} \: ,
\ee
where $V_0<0$, $R$ and $a_0$ are the depth, the radius and the 
diffuseness parameters of the well, respectively (given within the
following section), while the latter is

\be \label{coul}
   V_c(r)=\left\{\begin{array}{ll}
        \frac{Ze^2}{2R_c}\left(3-\frac{r^2}{R_c^2}\right) \: & r<R_c\\
        \frac{Ze^2}{r}                                   & r> R_c \: ,
        \end{array}\right.
\ee
for a uniformly charged sphere of radius $R_c$. The boundary 
conditions satisfied by these solutions are

\[
u_n(0)=0
\]
\[
\left. u_n(r)\right|_{r\to\infty}\to 0 \: .
\]

The eigenvalues of the WS potential well, which can not be found 
analytically, have been obtained by using the Nilsson-orbits computer
code developed by Hird \cite{hird}. A basis of harmonic oscillator 
eigenstates

\[
  R_{nl}(r)=\left[\frac{2\lambda^{3/2}n!}{\Gamma\left(n+l+\frac{3}{2}
            \right)}\right]^{\frac{1}{2}}(\lambda r^2)^{\frac{l}{2}}
            L_n^{l+\frac{1}{2}}(\lambda r^2)
            \exp{\left(-\frac{\lambda r^2}{2}\right)}
\]
has been used, with the condition for the best convergence of the WS 
states in the harmonic potential given by the relation \cite{hird}

\be \label{3.2}
   \lambda=0.15466\frac{\sqrt{|V_0|}}{R} \: ,
\ee
between the size parameter and the parameters of the well, where the 
energies are in MeV and the lengths in fm. Illustrative results of 
the numerical integration of Eq. (\ref{sch}) by using this method are 
shown in Fig. 1. 
\begin{figure}
\hspace{4.5cm}\epsffile{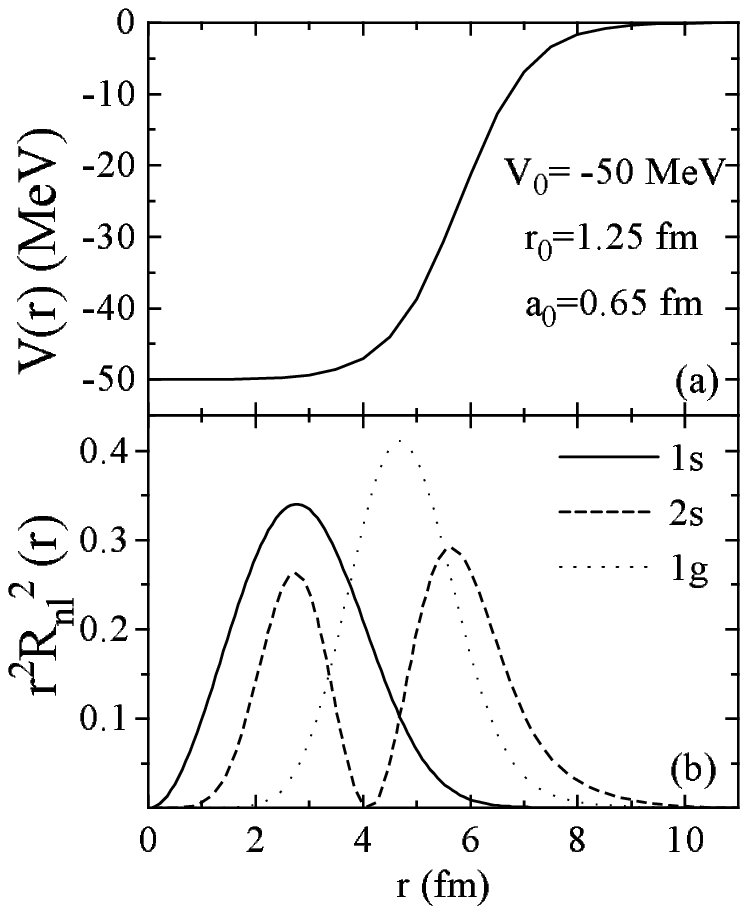}
\caption{
(a) The Woods-Saxon nuclear potential shape for the global parameter
set given, and (b) corresponding radial wave functions for neutrons.}
\label{fig:1}
\end{figure}

\subsection{Continuum-energy region ($\ve>0$)}

The methods used in this work for evaluation of the s.p.l. density 
within the bound-states and continuum regions, respectively, are
distinct. There are two linearly independent solutions for the radial 
Schr\"odinger equation, defined by various boundary conditions

\be \label{22.1}
\begin{array}{lcl}
   u_{lj}(r) \to \sin\left(kr-\frac{l\pi}{2}+\delta_{lj}\right) & 
                                        \mbox{when} & r\to\infty, \\
\end{array}
\ee
and
\be \label{22.2}
\begin{array}{lcl}
   \tilde u_{lj}(r)\to\cos\left(kr-\frac{l\pi}{2}+\delta_{lj}\right)& 
                                     \mbox{when} & r\to\infty \: . \\
\end{array}
\ee
Eq. (\ref{22.1}) is providing the regular solution \cite{new66,joac},
while Eqs. (\ref{22.1}) and (\ref{22.2}) can be used to obtain the 
irregular solution (Jost solution) $v_{lj}^{(\pm)}(r)$ satisfying the 
condition given by Eq. (\ref{21.2}),

\be \label{22.3}
    v_{lj}^{(\pm)}(r)=\tilde u_{lj}(r)\pm iu_{lj}(r) \: .
\ee

The wronskian of the two solutions is $r$-independent  (according to
the wronskian theorem) with the result $W_{lj}=-k$. Consequently, the 
imaginary part of Green's function becomes

\be \label{22.4}
\mbox{Im}\left(G_{lj}^{(+)}(r,r;\ve)\right)=-\frac{u^2_{lj}(r)}{k}\: .
\ee
Therefore, the $r$- and $\ve$-dependent s.p.l. density for $\ve>0$ is

\be \label{22.5}
g(\ve,r)=\frac{2m}{\pi\hbar^2}\frac{1}{k}\sum_{lj}(2j+1)u_{lj}^2(r)\: ,
\ee
i.e. it is given only by the regular solution of the radial 
Schr\"odinger equation corresponding to the potential well. 

The latest equations can be used for the calculation of the s.p.l. 
density for free particles too. The regular solution in this case is 
given by

\be \label{22.7}
     u^0_{lj}(r)=krj_l(kr) \: ,
\ee
where $j_l(kr)$ is the spherical Bessel function, regular in origin, 
and $g_0(\ve,r)$ becomes

\be \label{22.6}
     g_0(\ve,r)=\frac{2m}{\pi\hbar^2}kr^2\sum_{lj}(2j+1)j_l^2(kr) \: .
\ee
Since $2j+1=2(2l+1)$ in the general case of a single-particle 
potential which neglect the spin-orbit interaction, it results by 
using the relation \cite{abr}

\[ \sum_{l=0}^\infty(2l+1)j_l^2(x)=1, \]
that $g_0(\ve,r)$ is the same as given by Eq. (\ref{1.15}) except the 
spin degeneracy is taking into account

\[
g_0(\ve,r)=\frac{4m}{\pi\hbar^2}kr^2 \: .
\]
This result is an useful test for the correctness of the relation 
(\ref{22.5}) which has been in thiswork the starting point for the 
numerical calculation of the s.p.l. density in the continuum region.

The numerical solutions of the Schr\"odinger equation have been 
obtained by using the Cowell method \cite{meth,ber}. A modified 
version of the optical model code SCAT2 \cite{ber} has been used in 
this respect. The results of the numerical integration for neutrons 
is shown in Fig. 2.

The solutions are conveniently defined by the condition that they
are given asymptotically, e.g. for neutrons, by

\be \label{3.3}
    u_{lj}(r)=\sin\left(kr-\frac{l\pi}{2}+\delta_{lj}\right) \: \:
              \mbox{  when  }  r\to\infty \: ,
\ee
where $\delta_{lj}$ is the phase shift corresponding to the mean-field
potential. In particular, the phase shift can be used in the 
calculation of the differential cross section \cite{new66,joac} for 
the particles scattered on a finite-range potential (for the Coulomb 
potential can be also made). The $s$-wave phase shift for the WS 
potential well is shown as a function of energy in Fig. 2(b). 

\begin{figure}
\hspace{1.mm}\epsffile{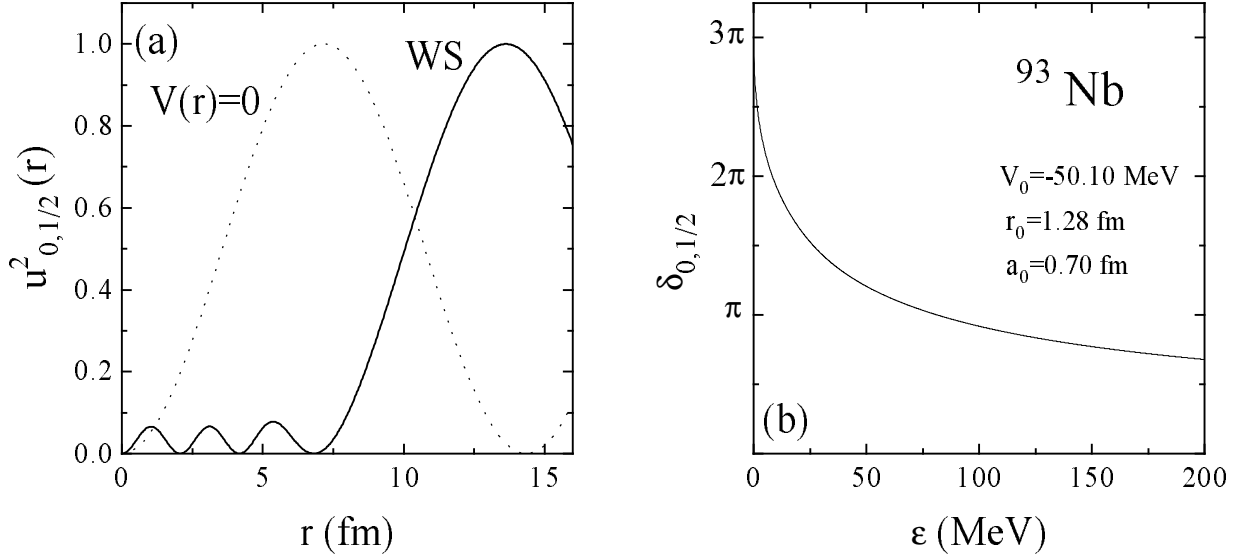}
\caption{
The s-wave radial solution of the Schr\"odinger equation in continuum,
for the Woods-Saxon potential (solid curve) as well as without it
(dotted curve), and (b) the corresponding s-wave neutron phase shift, 
for the nucleus $^{93}$Nb.}
\label{fig:2}
\end{figure}

\section{Model Parameters and Results }

The quantum-mechanical and TF calculations of the s.p.l. density for
neutrons within the nucleus $^{56}$Fe have been carried out by using 
the following potential wells. The respective parameters were 
determined by looking for the available experimental data on nuclear 
radii and separation energies to be reproduced by the corresponding 
single-particle Hamiltonian \cite{shl92}.

\subsection{Potential well parameters}


{\it (i) The harmonic-oscillator (HO) potential} as sample of an 
infinite potential well, but with a smooth surface and applicable 
only for $A\le$100, is given by

\be \label{ho}
    V^{HO}(r)=\frac{1}{2}m\omega^2r^2 \: ,
\ee
where the parameterisation \cite{shl92}

\be \label{3.31}
   \hbar\omega=\frac{45}{A^{1/3}}-\frac{25}{A^{2/3}} 
               \: \: \: \mbox{(MeV)}
\ee
has been adopted. On the other hand, the usual size parameter of the 
potential $\lambda=m\omega/\hbar$ has been replaced according to Eq.
(\ref{3.2}) for the basis of HO eigenstates which is used for the 
expansion of the WS potential eigenstates, for a better convergence 
of the numerical method \cite{hird}.

{\it (ii) The infinite square-well (SQ) potential} of the form

\be \label{sq}
   V^{SQ}(r)=\left\{ \begin{array}{ll}
                     V_0,    \:    & r<R      \\
                     \infty, \:    & r>R \: , \\
                     \end{array}\right.
\ee
has been used with the depth and radius parameters

\[
    V_0=-46+33\frac{N-Z}{A}t_3 \:\: \mbox{(MeV)},
\]
\[
      R=R_V=1.12A^{1/3}+1.6 \:\: \mbox{fm},
\]
Actually, it is the basis for the Fermi-gas model (FGM) of the 
nucleus. The respective eigenvalues are provided by the boundary 
condition for the single-particle wave functions to vanish at the 
nuclear surface ($r=R$) while the square-root energy dependence of
the s.p.l. density is given by, e.g., the semiclassical Thomas-Fermi 
formula (\ref{tf}).

{\it (iii) The finite square (FSQ) potential} of the form 

\be \label{fsq}
   V^{FSQ}(r)=\left\{\begin{array}{ll}
                     V_0 \: \: & r<R       \\
                     0   \: \: & r> R \: , \\
                     \end{array}\right.
\ee
has been used, with the depth \cite{shl92}

\[
  V_0=-47+33\frac{N-Z}{A}t_3 \: \: \:  \mbox{(MeV)} \: ,
\]
where $t_3$=1 for a neutron and -1 for a proton, while the radius has 
the same value as for the WS potential well. 

{\it (iv) The WS potential well} parameters within Eq. (\ref{ws}) as
given by Shlomo's global set are \cite{shl92}

\begin{eqnarray} \label{3.4}
  V_0 & = & -54+33\frac{N-Z}{A}t_3 \:\: \mbox{(MeV)} \nonumber \\
    R & = & \frac{R_V}{\left[1+(\pi a/R)^{1/3}\right]} \: ,
            \mbox{\hspace{1.5cm}}R_V=1.12A^{1/3}+1.0 \:\: \mbox{fm} \\ 
            \nonumber\\
    a & = & 0.70 \:\: \mbox{fm} \: , \nonumber 
\end{eqnarray}
where $R$ is determined by iteration. 

\subsection{Results and discussion}

The number of the bound single-neutron states with the energy less 
than $\ve$, i.e. $N(\ve)$ given by Eq. (\ref{num}), is shown in Fig. 
3(a) as a function of $\ve$ for the nucleus $^{40}$Ca, by using the 
WS potential within the both QM and TF methods. It is found similarly
to Shlomo \cite{shl92} that, in the case of the WS smooth potential,
the TF formula may be considered a good approximation of the 
quantum-mechanical results. Additionally, in Fig. 3(b) is shown the 
effect of the free-gas contribution subtracting, in the frame of the 
TF approximation.

The continuum effect on the s.p.l. density for neutrons is firstly 
shown within the TF method, in the case of the nucleus $^{56}$Fe. The
use of the infinite square-well provides the usual s.p.l. density  of
the Fermi-gas model. At the same time the respective finite well 
determines the continuum region of the spectrum where the 
consideration of the free-gas contribution by Eq. (\ref{subs}) 
is illustrated in Fig. 4(a). Next, the same effect in the TF 
approximation but using the realistic WS potential well is shown in
Fig. 4(b). It could be noted that, in spite of the quite distinct 
energy dependences due to the SQ and WS potentials wells, the
continuum components are rather similar following the proper 
consideration of the free-gas contribution.

The analysis of the continuum effect is completed by the 
quantum-mechanical calculation for the same neutron level density of 
the nucleus $^{56}$Fe in the Woods-Saxon potential well. The smooth
\begin{figure}
\hspace*{1mm}\epsffile{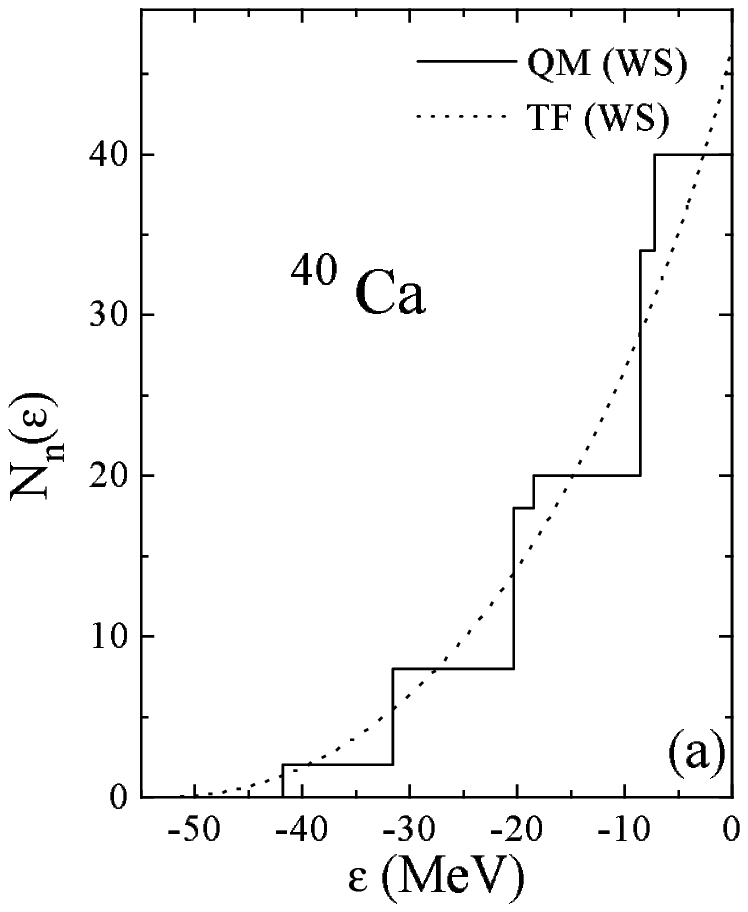}\epsffile{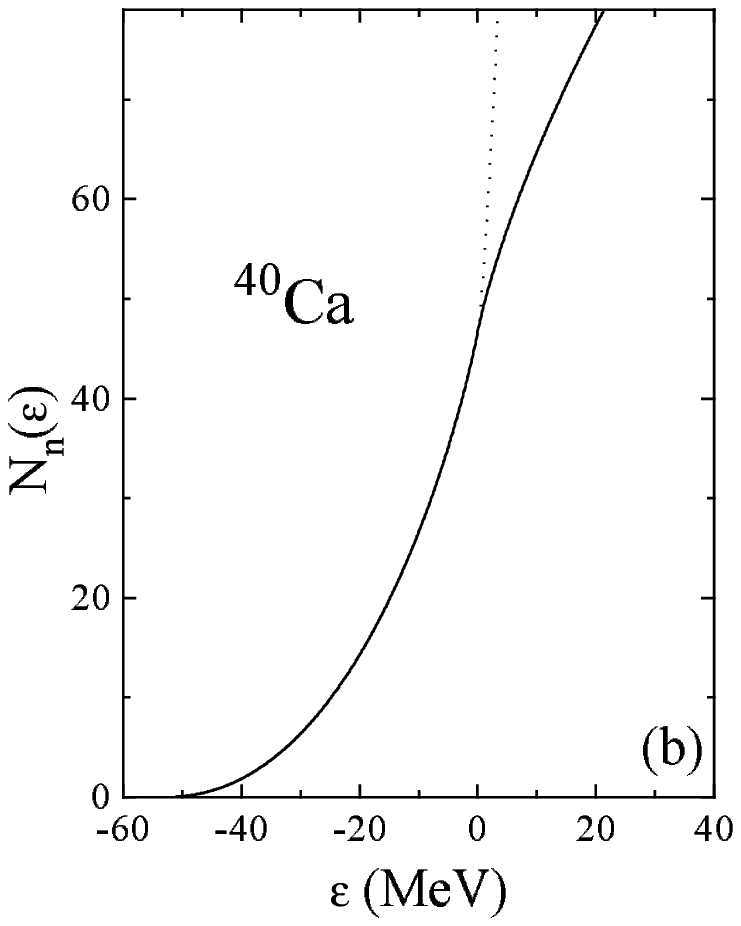}
\caption{
The number of single-neutron states $N(\ve)$ as a function of $\ve$ 
for the nucleus $^{40}$Ca and the Woods-Saxon potential, given by (a) 
the quantum-mechanical calculated single-particle energies 
(histogram), and the TF approximation (dotted curve), and (b) the 
latter method with (solid curve) and without (dotted curve) the 
free-gas contribution subtracted.}
\label{fig:3}
\end{figure}
\begin{figure}
\hspace*{1mm}\epsffile{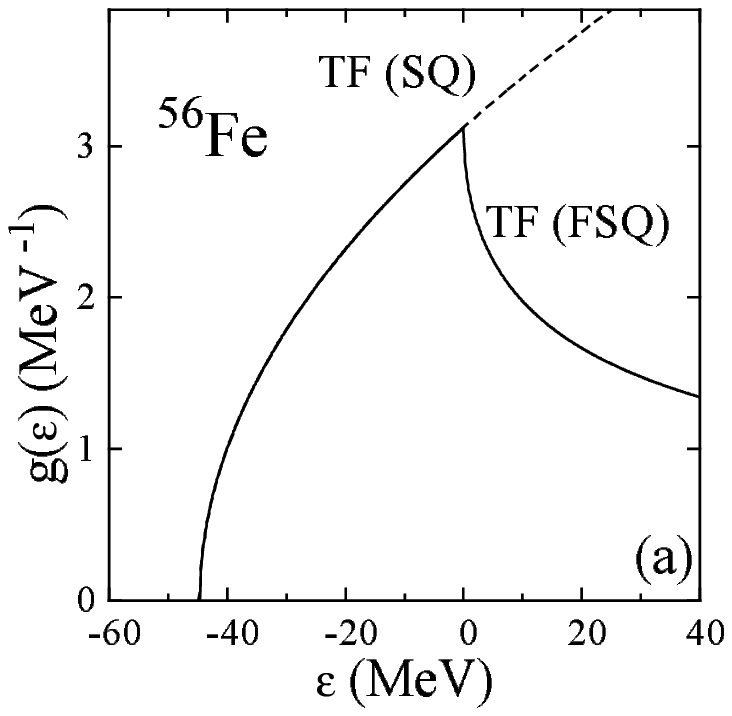}\epsffile{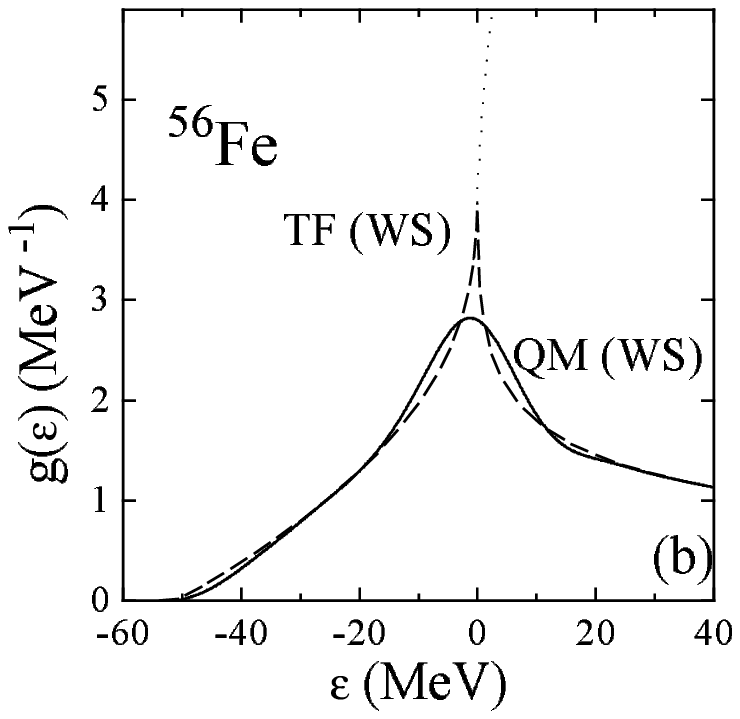}
\caption{
The s.p.l. density for neutrons of the nucleus $^{56}$Fe as a function 
of the single-particle energy, given by the TF approximation with (a)
the infinite square-well (dashed curve), finite square potential well 
(solid curve), and (b) the Woods-Saxon potential (dashed curve), as 
well as by the smoothed quantum-mechanical calculation with the same 
WS potential well (solid curve). The dotted curve correspond to the 
QM calculation carried out without subtraction of the free-gas 
contribution.}
\label{fig:4}
\end{figure}

\noindent
level density obtained by using the smooth parameter 
$\Gamma=1.2\hbar\omega$, where $\hbar\omega$ is given by Eq. 
(\ref{3.31}), and the order of Laguerre polynomial M=2, is shown in
Fig. 4(b) too. The comparison with the results of the TF 
approximation, which involves the same potential well, proves the 
suitability 
\begin{figure}
\hspace*{4.5cm}\epsffile{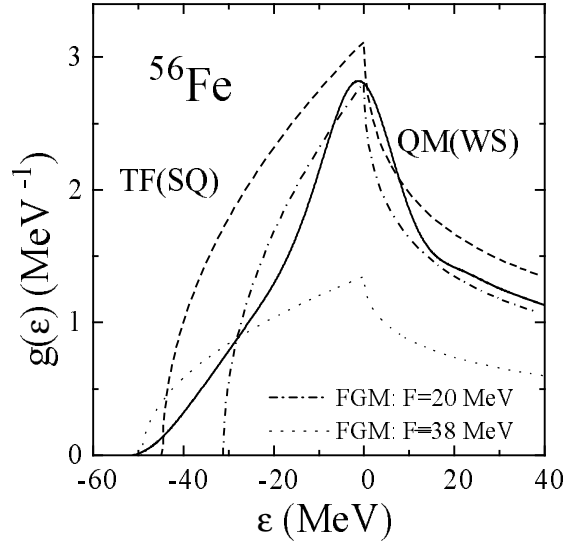}
\caption{
The same as Fig. 4, for the quantum-mechanical calculation with the
WS potential (solid curve), TF approximation with the FSQ potential 
well (dashed curve), and the FGM formula with the Fermi energy values
of $F$=38 MeV (dotted curve) and $\bar F$=20 MeV (dash-dotted curve).}
\label{fig:5}
\end{figure}
\noindent
of the latter similarly to that previously by Shlomo 
\cite{shl92}.

Actually, it results that the continuum component of the s.p.l.
density has rather close values within either exact QM calculations 
with the WS potential, or TF approximation with WS as well as FSQ
potential wells, provided that the free-gas contribution is 
subtracted. Moreover, a similar trend is obtained by means of the 
simple FGM formula for the s.p.l. density (e.g., \cite{kal}), if the 
continuum effect is taken into account (Fig. 5). The level of 
agreement between this phenomenological formula and the QM calculation 
could be improved for lower values of the Fermi energy $F$ which
is usually considered $\sim$38 MeV \cite{bohr}. It is thus even 
possible to reproduce the quantum-mechanical results by using the 
average Fermi energy $\bar F\sim$20 MeV which is determined by the
nuclear-surface localization of the first nucleon-nucleon collision
in preequilibrium reactions \cite{avr96,avr97,tri97}. 

\begin{figure}
\hspace{1mm}\epsffile{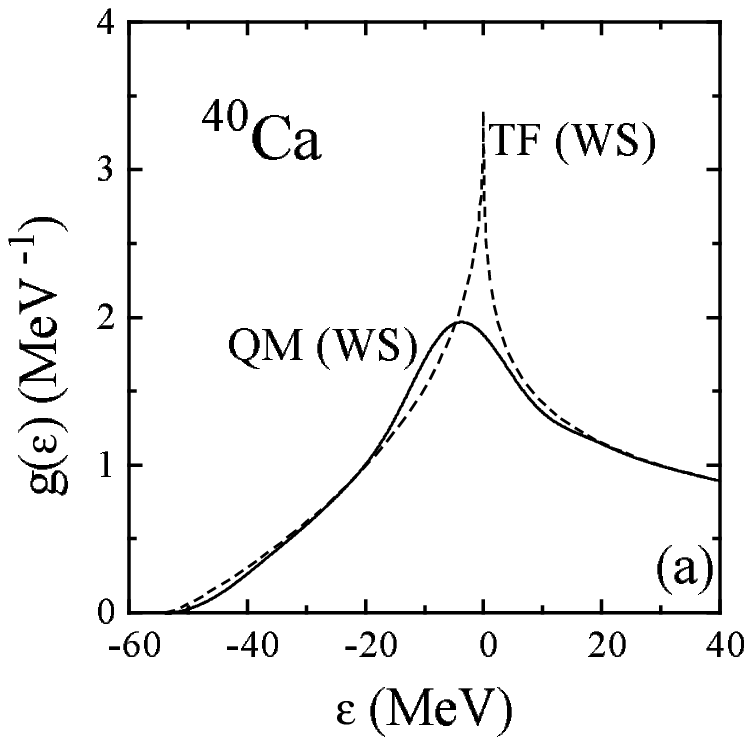}\epsffile{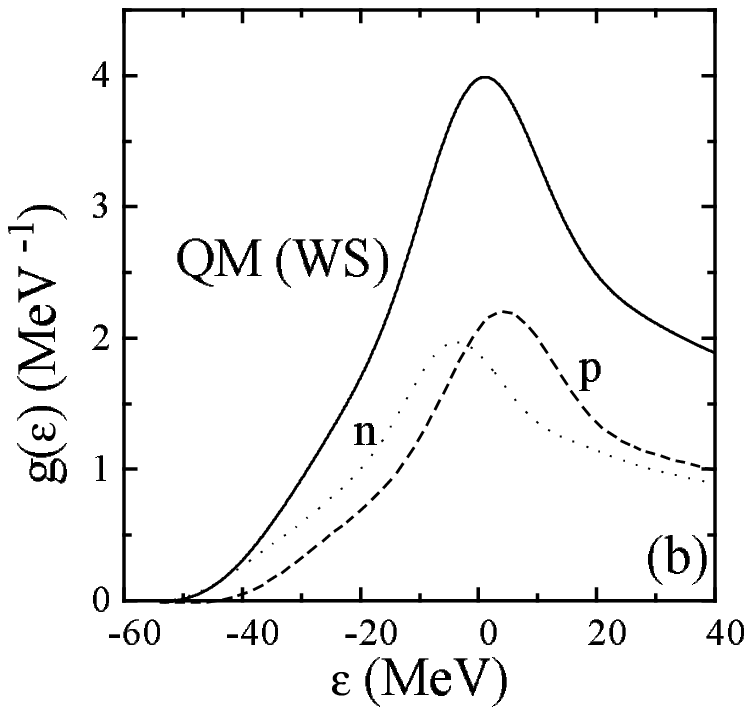}
\caption{
The smooth s.p.l. density for the nucleus $^{40}$Ca as a function of 
the single-particle energy, given by (a) quantum-mechanical (solid 
curve) and TF approximation (dashed curve), for neutrons, and (b) QM 
calculations for neutrons (dotted curve) as well as protons (dashed
curve) and their sum (solid curve).}
\label{fig:6}
\end{figure}

Finally, the smoothed results of the quantum-mechanical calculation 
for the s.p.l. density of neutrons as well as protons in the nucleus
$^{40}$Ca are shown in Fig. 6(b). In the proton case the calculation 
of the free-gas contribution has included the Coulomb interaction 
too. This case was chosen especially due to the equal numbers of 
neutrons and protons of $^{40}$Ca, so that it can be observed the 
effect of taking into account the both kinds of nucleons. Actually, 
the comparison of the exact QM and TF approximation is shown in Fig. 
5(a) only for neutrons, while next is given the QM calculated 
contributions of the both systems of nucleons as well as their sum.

\section{Conclusions}

A quantum-mechanical calculation of the single-particle level (s.p.l.) 
density $g(\varepsilon)$ is carried on by using the connection with 
the single-particle Green's function. The relation between the 
imaginary part of Green's function and single-particle wave functions 
is used separately for the discrete and continuous states. Within the 
bound-states region the imaginary part of the Green's function is 
calculated by using the wronskian theorem. The Green's function 
corresponding to the continuum is written by using the regular 
and Jost solutions of the radial Schr\"odinger equation. The smooth 
part of the rapidly fluctuating s.p.l. density is calculated by means 
of the Strutinski procedure. 

Recently Shlomo \cite{shl97} calculated the s.p.l. density by using 
the Green's function method, the phase-shift approach, and the TF 
approximation. The following different points could be underlined
between the present work and the Shlomo's analysis. First, the 
smearing procedure used by Shlomo \cite{shl92,shl97} is replaced 
with the exact calculation of the imaginary part of the Green's 
function carried out distinctively for the both bound and continuum 
regions of the spectrum. Second, Shlomo has considered the nucleus 
contained in a large but finite box within both the Green's function 
method and the phase-shift approach. This condition is not required 
in the present work. Third, we have taken into account the case of 
the protons too, where the Coulomb interaction has to be also 
included in the contribution of the free gas.

Finally, it results that the continuum component of the s.p.l.
density has rather close values within either exact QM calculations 
with the WS potential, or TF approximation with WS as well as FSQ
potential wells, provided that the free-gas contribution is 
subtracted. A similar trend is obtained by means of the simple FGM 
formula for the s.p.l. density, if the continuum effect is taken into 
account. The level of agreement between this phenomenological formula 
and the QM calculation could be improved for lower values of the Fermi 
energy $F$ which is usually considered $\sim$38 MeV \cite{bohr}. It is 
thus even possible to reproduce the quantum-mechanical results by 
using the average Fermi energy $\bar F\sim$20 MeV 
\cite{avr96,avr97,tri97}. Additional analysis is yet required by, 
e.g., adoption of realistic mean-field potentials energy dependent 
\cite{shl97}, and systematic calculations of s.p.l. density including
the continuum effect.

\acknowledgments
The author is grateful to Prof. Dr. Gh. Ciobanu and Dr. V. Avrigeanu 
for making possible this work and the final form of the paper,  
Prof. Dr. V. Florescu and  Dr. M. Avrigeanu for valuable discussions,
and Dr. M. Mirea for making available the Nilsson-orbits code. The
useful suggestions and correspondence from Dr. S. Shlomo are much
acknowledged too. This work was carried out under the Romanian 
Ministry of Research and Technology Contract No. 34/A13.


\begin{thebibliography}{99}
\bibitem{shl92} S. Shlomo, Nucl. Phys. {\bf A539}, 17 (1992)
\bibitem{shl96} Ye. A. Bogila, V. M. Kolomietz,
A. I. Sanzhur, and S. Shlomo, Phys. Rev. {\bf C53}, 855 (1996)
\bibitem{shl97} S. Shlomo, V. M. Kolomietz, and H. Dejbakhsh, Phys.
Rev. {\bf C55}, 1972 (1997)
\bibitem{ross}  C. K. Ross and R. K. Bhaduri, Nucl. Phys.
{\bf A188}, 566 (1972)
\bibitem{new66} R. Newton, {\it Scattering Theory of Waves and 
Particles}, McGraw-Hill, New York, 1966.
\bibitem{goriely96}  S. Goriely, Nucl. Phys. {\bf A605}, 28 (1996)
\bibitem{bal70} R. Balian, and C. Bloch, Ann. Phys.(NY)
{\bf 60}, 401 (1970)
\bibitem{bra} M. Brack and H. C. Pauli, Nucl. Phys. {\bf A207},
401 (1973)
\bibitem{abr} M. Abramowitz and I. Segun, {\it Handbook of 
  mathematical functions}, 
  7th edition, Dover Publication, New York.
\bibitem{ring} P. Ring and P. Schuck, {\it The Nuclear Many-Body 
Problem}, Spriger, Berlin, 1980, Ch. 13.
\bibitem{joac} C. J. Joachain, {\it Quantum Collision Theory}, 3-rd 
edition, North-Holland, Amsterdam, 1983.
\bibitem{hird} B. Hird, Comp. Phys. Com., {\bf 6}, 30 (1973)
\bibitem{meth} B. Alder (editor), {\it Methods in Computational 
Physics}, Academic Press, New York, 1966.
\bibitem{ber} O. Bersillon, {\it Un programme de modele optique 
spherique}, CEN-Bruyeres-le-Chatel, Note CEA-N-2227, 1981.
\bibitem{kal} C. Kalbach, Phys. Rev. {\bf C32}, 1157 (1985)
\bibitem{bohr} A. Bohr, B. R. Mottelson, {\it Nuclear Structure},
Benjamin, New York, 1969, vol. 1.
\bibitem{avr96} M. Avrigeanu, A. Harangozo, V. Avrigeanu, 
and A. N. Antonov, Phys. Rev. {\bf C54}, 2538 (1996)
\bibitem{avr97} M. Avrigeanu, A. Harangozo, V. Avrigeanu, 
and A. N. Antonov, Phys. Rev. {\bf C56}, 1633 (1997)
\bibitem{tri97} M. Avrigeanu, A. Harangozo, I. \c Ste\c tcu, and 
V. Avrigeanu, Proc. Nuclear Data for Science and Technology, Trieste,
1997 ({\it in press}).
\end{thebibliography}
\end{document}